\documentclass{article}
\usepackage[hypertex]{hyperref}
\usepackage{amsfonts}
\usepackage[OT2,T1]{fontenc}
\newcommand\textcyr[1]{{\fontencoding{OT2}\fontfamily{wncyr}\selectfont #1}}

\title{The April First Phenomenon}
\author{George Svetlichny\footnote{Departamento de Matem\'atica, Pontif\'{\i}cia Universidade Cat\'olica, Rio de Janeiro, Brazil \newline
svetlich@mat.puc-rio.br \hfill \url{http://www.mat.puc-rio.br/\~svetlich}}}
\date{April first 2014}
\begin{document}
\maketitle
\begin{abstract}
A true quantum reason for why people fib on  April first.
\end{abstract}

\section{The truth\protect\ldots}
One of the least understandable of human phenomena is the propensity to fib on  April first. How is it that an activity so reprehensible on other days of the year is so readily accepted on this one singular day? Many theories have been put forth about this \cite{svet:jbbt3.14,tem:vov271.8281,perr:apoc} usually of a sociological type. However, sociology alone cannot explain this. This is because even scientists, for whom the truth is the highest virtue, succumb to this failing. One can attest to this fact by numerous text published by such reputable venues as Scientific American or arXiv. Physicist, for whom truth and precise language is such an uncompromising commitment as to make them, in gentle terms, the least tolerable of the science workers\cite{bbt}, and who would not jeopardise their professional standing and careers in such a manner, nevertheless engage in this practice. This observation shows that we have to seek deeper causes for the phenomenon, and obviously, only quantum physics can supply this.

So one comes to this paradox: physicists are perceived to sometimes lie in reputable scientific venues and yet they cannot lie. The obvious conclusion is that they are not lying but telling the truth, and it is the nature of the universal physical state on each  April first that somehow makes us believe, after the fact, that lies have been perpetrated. As already mentioned, quantum physics has an obvious explanation as to how this can happen. In the Everett many-worlds interpretation of quantum mechanics\cite{slide} there are the so called \emph{Maverick Universes} in which the ordinary laws of physics can break down because quantum probabilities don't follow the usual Born rule. It must surely be that on each April first we enter a maverick universe and so what appears to be fibs are in fact solid truths \emph{in the current unverse}. This settles that. April Fools' Day proves the truth of the Everett picture.

But why is it that it is precisely on one day of the year and not on any other day that we \emph{slide} from one universe to another? Cosmic alignment of this planet with whatever unknown thing is out there cannot be the cause. It could not be a close whatever for our sun is speeding through our galaxy and so alignment days would not repeat. With a distant whatever \emph{all} days would be like April Fools' Day, and they're not. With intermediate distances the singular day would slowly slide trough the calendar but it's been steady for centuries. No, one needs a local explanation, and,  as expected, quantum physics provides the answer. As April first nears, many people on the planet perform a quantum measurement on their friends and colleagues by telling fibs to see if their friends and colleagues accept them or not. The latter, exercising their free will, go along so as to humor the fibbers. Free will does not have to follow the Born rule, so the unverse slowly gets pushed into a maverick state. As fibbing continues one enters into a quantum Zeno process by which the state of the universe freezes as a maverick. Physicists are notoriously detached and distracted\cite{bbt}, don't notice the goings-on, and  working in a  maverick universe perform experiments and calculations that give what would otherwise be totally absurd and contradictory results but which on that day are truly true. Unwittingly they publish. This settles that.

That it is April first that people in general put the universe through a quantum Zeno process is pure coincidence; the practice started somehow, caught on, and became a tradition.\footnote{Look up ``April Fools' Day" in Wikipedia, if you're so inclined, but be sure to check the dates of writing.}  This is sociology.\footnote{Yes, unfortunately some sociology had to creep in. Sorry.} It could have been any other day of the year. This is spontaneous symmetry breaking.

\section{\protect\ldots shall set you free}

There is a universally unnoticed side effect of the April First phenomenon. Quantum probabilities that do not follow the Born rule allow for superluminal communication and so also for all of its collateral advantages.

So, on April first, if you are very clever, you can instantly communicate with distant galaxies, solve any hard computation problem in polynomial time, send messages to the past, move things with your mind, generate any amount of energy, and maybe even become immortal. But only on that day.

Carpe Diem!


\end{document}